\documentclass[twocolumn,prl,twocolumn,superscriptaddress]{revtex4-1}
\usepackage{amsmath,amssymb,mathrsfs}
\usepackage{natbib}
\usepackage{subfigure}
\usepackage{tabularx}
\usepackage{epsfig}
\usepackage{longtable}
\usepackage{amsfonts}
\usepackage{rotating}
\usepackage{subfigure}
\usepackage{amsmath}
\usepackage{comment}
\usepackage{bbold} 
\usepackage{hhline}

\def\be{\begin{equation}}
\def\ee{\end{equation}}
\def\bea{\begin{eqnarray}}
\def\eea{\end{eqnarray}}
\def\nn{\nonumber}

\def\up{\uparrow} 
\def\down{\downarrow}
\usepackage{wordlike}

\usepackage[unicode=true,bookmarks=true,bookmarksnumbered=false,bookmarksopen=false,breaklinks=false,pdfborder={0 0 1},backref=false,colorlinks=true]{hyperref}

\hypersetup{linkcolor=magenta,urlcolor=blue,citecolor=blue,pdfstartview={FitH},hyperfootnotes=false,unicode=true}

\def\be{\begin{equation}}
\def\ee{\end{equation}}
\def\bea{\begin{eqnarray}}
\def\eea{\end{eqnarray}}
\def\nn{\nonumber}

\def\up{\uparrow}
\def\down{\downarrow}
\def\psihat{\hat{\psi}} 
\def\CB{\hat{\cal B}}

\begin{document}
\title{Chiral Induced Spin Selectivity as a Spontaneous Intertwined Order}
\author{Xiaopeng Li}
\email{xiaopeng\_li@fudan.edu.cn}
\affiliation{State Key Laboratory of Surface Physics, Institute of Nanoelectronics and Quantum Computing, and Department of Physics, Fudan University, Shanghai 200438, China}
\affiliation{Shanghai Qi Zhi Institute, AI Tower, Xuhui District, Shanghai 200232, China}
\author{Jue Nan} 
\affiliation{State Key Laboratory of Surface Physics, Institute of Nanoelectronics and Quantum Computing, and Department of Physics, Fudan University, Shanghai 200433, China}
\author{Xiangcheng Pan} 
\affiliation{State Key Laboratory of Molecular Engineering of Polymers, Department of Macromolecular Science, 
Fudan University, Shanghai 200438, China}

\begin{abstract}
{Chiral induced spin selectivity (CISS) describes efficient spin filtering by chiral molecules. This phenomenon has led  to nanoscale manipulation of quantum spins with promising applications to spintronics and quantum computing, since its discovery nearly two decades ago. However, its underlying mechanism still remains mysterious for the required spin-orbit interaction (SOI) strength is unexpectedly large. Here we report a multi-orbital theory for CISS, where an effective SOI emerges from spontaneous formation of electron-hole pairing caused by many-body correlation. 
This mechanism produces a strong SOI reaching the energy scale of room temperature, which could support the large spin polarization observed in CISS. One central ingredient  of our theory is the Wannier functions of the valence and conduction bands correspond  respectively to one- and two-dimensional representation of the spatial rotation symmetry  around the molecule elongation direction. The induced SOI strength is found to decrease when the band gap increases. Our theory may provide important guidance for searching other molecules with CISS effects. 
}

\end{abstract}

\date{\today}

\maketitle

{\it Introduction.---}
Atomic spin-orbit coupling is a relativistic quantum effect that originates from the fundamental quantum electrodynamics of electrons orbiting around the nucleus. It is established that heavier atoms tend to have stronger spin-orbit couplings as the effective coupling strength increases with the atomic number $Z$ as  $Z^4$~\cite{1963_Herman_PRL,2014_Balents_ARCMP}.  Consequently, material research aiming for strong spin-orbit effects has been mainly focusing on materials composed of heavy atoms~\cite{2014_Balents_ARCMP}. 
 
 \begin{figure}
\begin{center}
\includegraphics[width=.9\linewidth]{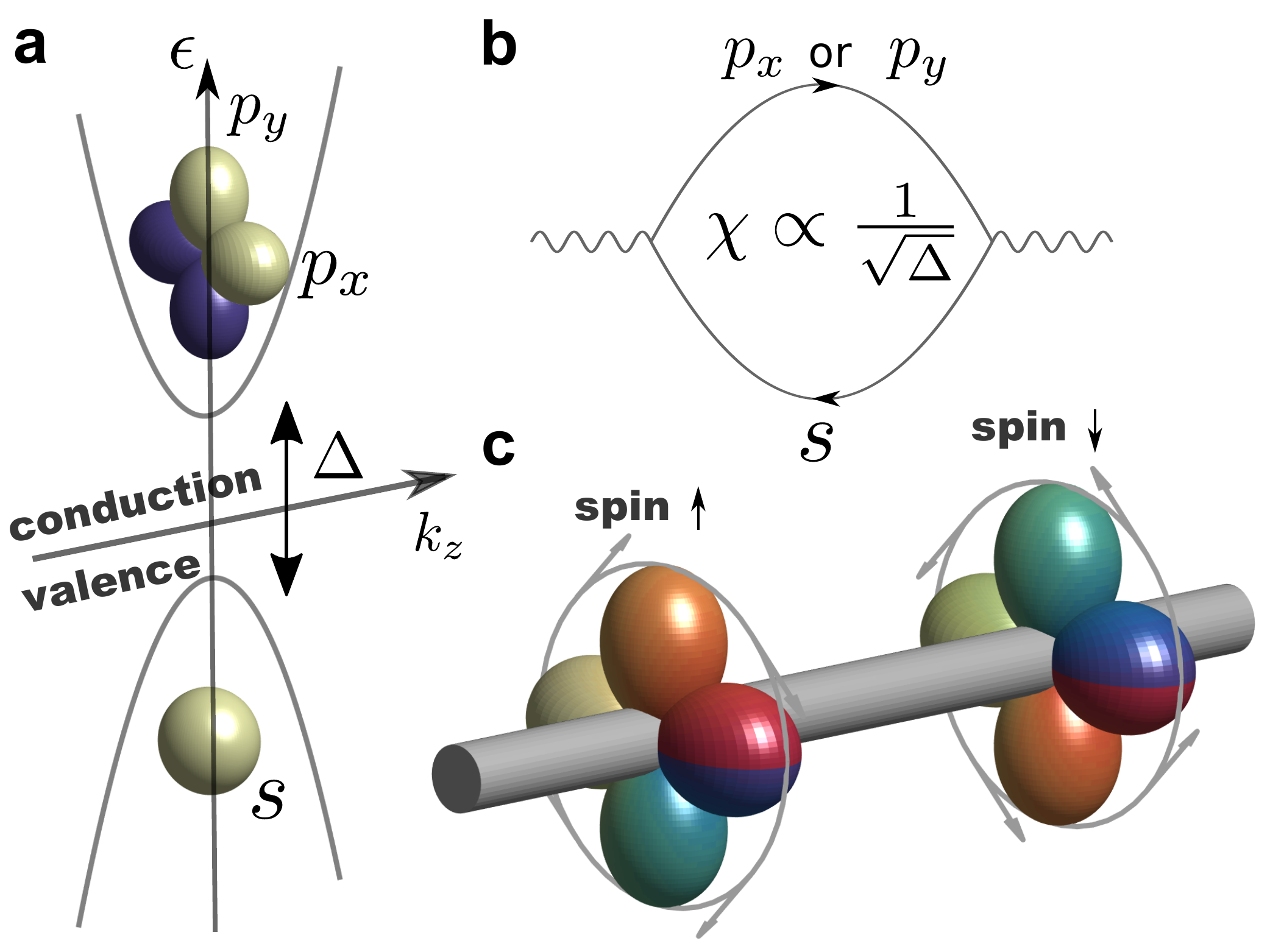}
\end{center}
\caption{{\bf Electron-hole paring in the three-orbital model.} {\bf a}, Illustration of the three orbital ($s$, $p_x$, and $p_y$) model. The valence and conduction bands have $s$-and $p$-orbital character, respectively. The band gap is $\Delta$. {\bf b}, The Feynman diagram that leads to the divergent susceptibility towards forming electron-hole paring across the valence and conduction bands. {\bf c}, Sketch of the spin-orbit intertwined order with the spin and angular-momentum of the electron-hole pair spontaneously coupled.  }
\label{fig:f1}
\end{figure}

However, a surprisingly large spin-orbit effect has been found in chiral organic and biological molecules mainly composed of carbon atoms in the study of chiral induced spin selectivity~\cite{1999_Ray_Science}, which has received enormous research efforts in the last decade~\cite{2019_Naaman_NatureReview,2012_Naaman_JPCL,2011_Gohler_Science,2011_XieNanoLett,2013_Mishra_PNAS,2013_Paltiel_NatComm,2014_Paltiel_NanoLetter,2016_Naaman_NatComm,2016_Naaman_Advanced,2017_Naaman_NatComm,2017_Aragones_Small,2018_Naaman_Advanced,2018_Zacharias_JPCL,2019_Naaman_JPCL,2019_Naaman_Small}. 
This fascinating phenomenon has been observed in a broad range of  chiral molecules, from DNA~\cite{1999_Ray_Science,2011_Gohler_Science} and protein~\cite{2013_Mishra_PNAS}, to  $\alpha$-helical peptides~\cite{2017_Aragones_Small,2016_Naaman_NatComm} and helicene~\cite{2016_Naaman_Advanced,2018_Zacharias_JPCL}, using a number of different experimental setups such as photoelectron transmission~\cite{1999_Ray_Science,2011_Gohler_Science}, transport~\cite{2017_Aragones_Small,2011_XieNanoLett,2016_Naaman_Advanced}, and electrochemistry measurements~\cite{2015_Naaman_Advanced}. 
It has far reaching implications in the fundamental understanding of important biological processes such as protein-folding and biorecognition~\cite{2019_Naaman_NatureReview}.  
The experimental observations imply a generic underlying mechanism of spin filtering by chiral molecules, that demands a theoretical explanation. 
Previous theoretical studies have shown that a sufficiently strong SOI is required for tight-binding models to reproduce the experimental features~\cite{2012_Guo_PRL,2012_Gutierrez_PRB,2015_Naaman_Review,fransson2019chirality}, 
although the intrinsic spin orbit interaction in these molecules mainly containing carbon atoms is too weak to accommodate the observed large spin polarization, for example up to 60\% using chiral peptide molecules~\cite{2017_Aragones_Small}.   
The essential question that remains outstanding is how the unexpectedly strong SOI emerges~\cite{2018_Rebergen_CIIS} beyond the conventional consideration of the quantum electrodynamics.

In the study of atomic Bose-Einstein condensates, a quantum fluctuation enabled spin-momentum intertwined order has been proposed in theory~\cite{2014_Li_NatComm} to explain the observed spin-momentum locking in a hexagonal optical lattice without bare SOI~\cite{2012_Sengstock_NatPhys}. This mechanism has been further generalized to a multi-orbital Bose-Einstein condensate, where a spontaneous spin angular-momentum intertwined order is shown to occur in a meta-stable state for spinor Bosons residing on excited bands of a square lattice~\cite{2018_Li_PRL}. These theoretical studies are inspiring, suggesting that spin-orbit coupling could emerge from many-body correlation even in complete absence of single-particle SOI, although the  bosonic theories do not apply to electrons bearing CISS in chiral molecules.

In this work, we assume a rotation symmetry around the molecule elongation direction (to be refereed as $z$ below), and consider a setup where electronic valence and conduction bands correspond to a one- and two-dimensional representation of the symmetry group, respectively. The Wannier functions then have $s$ and $p$-orbital character (Fig.~\ref{fig:f1}). Through field theory analysis and renormalization group calculation, we show the electron correlation causes a strong instability towards forming electron-hole pairs that spontaneously break the spin SU(2) and reflection symmetries with preservation of time-reversal. This spontaneous spin-orbit intertwined order gives rise to a strong SOI having a first-quantization form 
\be 
\lambda_{\rm so} \hat{ \sigma}_z \otimes \hat{L}_z/2 , 
\ee 
with $\hat{\sigma}_z$ the spin Pauli-z operator and $\hat{L}_z$ the angular momentum operator for $p$-orbitals in the conduction band. 
Considering a band gap $\Delta$, $s$- and $p$-orbital tunnelings, $t_s$ and $t_p$, and their interactions including density-density interaction $U$, Hund's rule coupling $J$, and Josephson coupling $J'$~\cite{supplement}, 
the induced SOI strength is given as 
\be 
\lambda_{\rm so} = \Delta /2 
- \sqrt{ \Delta ^2 /4 + (U-J-J')^2 |\phi|^2 /2 } , 
\label{eq:SOIstrength} 
\ee 
with $|\phi|$ the amplitude of the spin-orbit intertwined order parameter estimated to be 
$|\phi| = 0.25\times (U-J-J')/(t_s + t_p) $ in the small band gap limit. 
{
Taking an example of $t_s + t_p = 2$ eV, $U-J-J' = 1$ eV and $\Delta = 0$---the parameter choice corresponds to an interaction estimate~\cite{supplement} for a right-handed peptide $3_{10}$ helix with a triplet-paired band structure~\cite{2020_Yan_CISS}---the induced SOI reaches $0.1$ eV, which suffices for modeling the unexpectedly large CISS effects in chiral molecules that persist upto room temperature~\cite{2018_Rebergen_CIIS}.}  
The strength of the induced SOI decreases when the band gap is increased. 
We expect this result would strongly contribute to the  understanding of spin-selective processes in biology~\cite{2019_Naaman_NatureReview}. 


\begin{table*} [htp]
\begin{tabular} {| c | c | c | c | c|} 
\hhline{=====} 
$j_s$, $m_s$, $m_l$ 				&Operators			& SU(2) 			&Parity		& TRS\\ \hline \hline
$0, 0, 0$	
								& $\begin{array}{c} 
											\CB_{0,0,0;0}, ( \CB_{0, 0, 0; +1}  + \CB_{0,0,0; -1})/\sqrt{2} \\ 
											(\CB_{0, 0, 0; +1}- \CB_{0,0,0; -1})/\sqrt{2}  
									\end{array}$ 
													&Singlet  			&Even 
																				& $\begin{array}{c} 
																				{\rm Even} \\ 
																				{\rm Odd} 
																				\end{array} $ 	
 \\ \hline 
 $0, 0, 1$ 
								& $\begin{array}{c} 
								(\CB_{0, 0, 1; 0} + \CB_{0, 0, 1; -1} )/\sqrt{2}  \\ 
								(\CB_{0, 0, 1; 0}- \CB_{0, 0, 1; -1})/\sqrt{2} 
								\end{array} $ 			
													&Singlet 			&Odd
														 						&$\begin{array}{c} 
																				{\rm Even} \\ 
																				{\rm Odd} 
																				\end{array} $ 
\\ \hline 
$0, 0, 2$ 
											&$\CB_{0,0,2;-1}$	
													&Singlet 		&Even	&Even 
\\ \hline 
$1, m_s\in\left\{0,\pm 1\right\}, 0$ 									
								&  $\begin{array}{c} 
								\CB_{1,m_s,0;0}, ( \CB_{1, m_s, 0; +1}  + \CB_{1, m_s,0; -1})/\sqrt{2} \\ 
								(\CB_{1, m_s, 0; +1}- \CB_{1,m_s,0; -1})/\sqrt{2}  
								\end{array}$ 	
													& Triplet 			&Even 
																				&$\begin{array}{c} 
																				{\rm Odd} \\ 
																				{\rm Even} 
																				\end{array} $
\\ \hline 
$1, m_s\in\left\{0,\pm 1\right\}, 1$ 									
								& $\begin{array}{c} 
								(\CB_{1, m_s, 1; 0} + \CB_{1, m_s, 1; -1} )/\sqrt{2}  \\ 
								(\CB_{1, m_s, 1; 0}- \CB_{1, m_s, 1; -1})/\sqrt{2} 
								\end{array} $ 		
													& Triplet			&Odd
																				&$\begin{array}{c} 
																				{\rm Odd} \\ 
																				{\rm Even} 
																				\end{array} $ 
\\ \hline 
$1, m_s\in\left\{0,\pm 1\right\}, 2$ 		& $\CB_{1, m_s,2;-1}$	
													& Triplet 			&Even		&Odd\\ 
\hhline{=====}  
\end{tabular} 
\caption{{\bf Symmetry properties of electron-hole pairings.} 
We have introduced quantum numbers $j_s$ and $m_s$ according to the spin SU(2) symmetry, with $j_s$ equal to $0$ and $1$ labeling singlet and triplets. 
Under a spatial rotation around the $z$ direction by an angle $\theta$, the operators transform as $B_{j_s, m_s, m_l; q} \to B_{j_s, m_s, m_l; q} e^{im_l \theta}$, determined by the quantum number $m_l$. 
Under time-reversal symmetry (TRS) transformation, we have $ B_{j_s, m_s, m_l; q}  \to (-)^{j_s + 1} (-)^{m_l + m_s} B_{j_s, -m_s, -m_l; -q} $. 
The even/odd sign of TRS listed here is determined for an operator $\hat{\cal O}$ according to whether its corresponding Hermitian observable $h\hat{\cal O} + h^*\hat{\cal O}^\dag$ (with $h$ an arbitrary complex number) is TRS even or odd.  The operators with $m_l = 0, -1, -2$ are not listed here due to the constraint that $ B_{j_s, m_s, m_l; q} ^\dag = (-)^{m_l+m_s + 1} B_{j_s, -m_s, -m_l; q+m_l} $. 
} 
\label{tab:t1} 
\end{table*}

\medskip 
Our theory starts from a field theory description of the three-orbital system (Fig.~\ref{fig:f1}), 
\bea 
\hat{H}_0 &=& \int d z \sum_{\nu, \alpha} P_\nu \hat{ \psi}_{\nu \alpha} ^\dag (z)  \left[\frac{\hbar^2\partial_z ^2   } {2m_\nu }-\frac{\Delta} {2}\right]  \hat{\psi}_{\nu \alpha} (z). 
\label{eq:fieldHam} 
\eea 
Here $\nu=s$, $p_x$, or $p_y$ index the orbitals, and $\alpha$ the spin degrees of freedom; $P_\nu$ represents the parity, i.e., $+$ and $-$ for $s$- and $p$-orbitals, respectively; $m_s$ and $m_{p_x} = m_{p_y} \equiv m_p$ the effective mass associated with the motional dynamics of  the valence and conduction bands along the molecular elongation direction. The field operators $\hat{\psi}_{\nu \sigma}$ incorporates the low energy degrees of freedom of  electrons moving in a molecule near the band edge. 
{ 
For example, in modeling a right-handed peptide helix with a triplet-paired band structure~\cite{2020_Yan_CISS}, our three-orbital model corresponds to an effective description of electronic properties of the molecule, with $s$ and $p$-orbitals representing the quantized transverse modes. 
Considering a carbon atomic chain, our theory corresponds to  the Fermi energy lying in between the $sp$-hybridized $\sigma^*$-bond and $\pi$-bond. 
} 
We remark that the $s$-orbital in our model may represent a $p_z$ orbital in the molecule as well, which obeys the same symmetry under the spatial rotation around the $z$ direction.

An immediate consequence of the field theory is that it develops strong susceptibility towards electron-hole pairing at low temperature even without interaction. This is described by a response function,  
$
\chi_{sp} ^0=\partial_h  \langle \hat{\psi}_{s \alpha} ^\dag \hat{\psi}_{p\alpha'} + H.c. \rangle = \sqrt{( m_s ^{-1} + m_p^{-1} )/{2\Delta} }, 
$ 
considering a fictitious infinitesimal perturbation $ \Delta H = -h \int d z  \left[\hat{\psi}_{s \alpha} ^\dag \hat{\psi}_{p\alpha'} + H.c.\right]$. 
This response has a divergent behavior $1/\sqrt{\Delta}$  for a small band gap, which is caused by the interplay of the logarithmic divergence of the Feynman diagram (Fig.~\ref{fig:f1}) and the van Hove singularity in one dimensional density of states.  
In order to have spin-orbit coupled effect from the divergent electron-hole pairing, the idea here is to form a quantum superposition between the $s$-orbital and the finite angular momentum $p_x \pm ip_y$ state that preserves time-reversal symmetry. This would then provide the required effective spin-orbit coupling.  To form an orbital superposition in this channel, we need to assume that  the Wannier functions of the valence and conduction bands correspond  respectively to one- and two-dimensional representation of the spatial rotation symmetry  around the molecule elongation direction. Such physics does not occur in the previous study of correlation effects in a single-band setting~\cite{fransson2019chirality}.

The divergence in the response function indicates important many-body effects in the system. 
Having spin SU(2) and spatial rotation symmetries,  the three interaction terms including density-density interaction $U$,   Hund's rule coupling $J$, and Josephson coupling $J'$~\cite{supplement}
are the only allowed local interactions between the $s$- and $p$-orbitals.   Due to the multi-orbital complexity of our model, there are $21$ independent channels in the particle-hole pairing function, $G_{\nu \alpha, \nu' \alpha'} = \langle \hat{\psi}^\dag _{\nu \alpha} \hat{\psi}_{ \nu'\alpha'} \rangle$. 
According to  symmetry properties, we group all the particle-hole pairings into the following channels, 
\bea 
\hat{\cal B}_{j_s, m_s, m_l; q} =  (-1)^{q+1}  \sum_{\alpha \alpha'}  i^{2\alpha } C_{\frac{1}{2} \frac{1}{2}} (-\alpha, \alpha'| j_s, m_s) \hat{\psi}_{q \alpha} ^\dag \hat{\psi}_{q+m_l \alpha'}  \nn\\
\label{eq:Bop} 
\eea  
where the $C_{\frac{1}{2} \frac{1}{2}} $ matrix contains Clebsch-Gordon coefficients, 
and $\hat{\psi}_{q\alpha}$ is  a field operator in angular momentum basis defined by 
$[\hat{\psi}_\pm\equiv \mp (\hat {\psi}_x \pm i\hat{\psi}_y)/\sqrt{2}$, $\hat{\psi}_{0} \equiv \hat{\psi}_{s } ]$.  
The symmetry properties of these operators are listed in Table~\ref{tab:t1}. 


\begin{figure*}[htp]
\centerline{\includegraphics[angle=0,width=.8\textwidth]{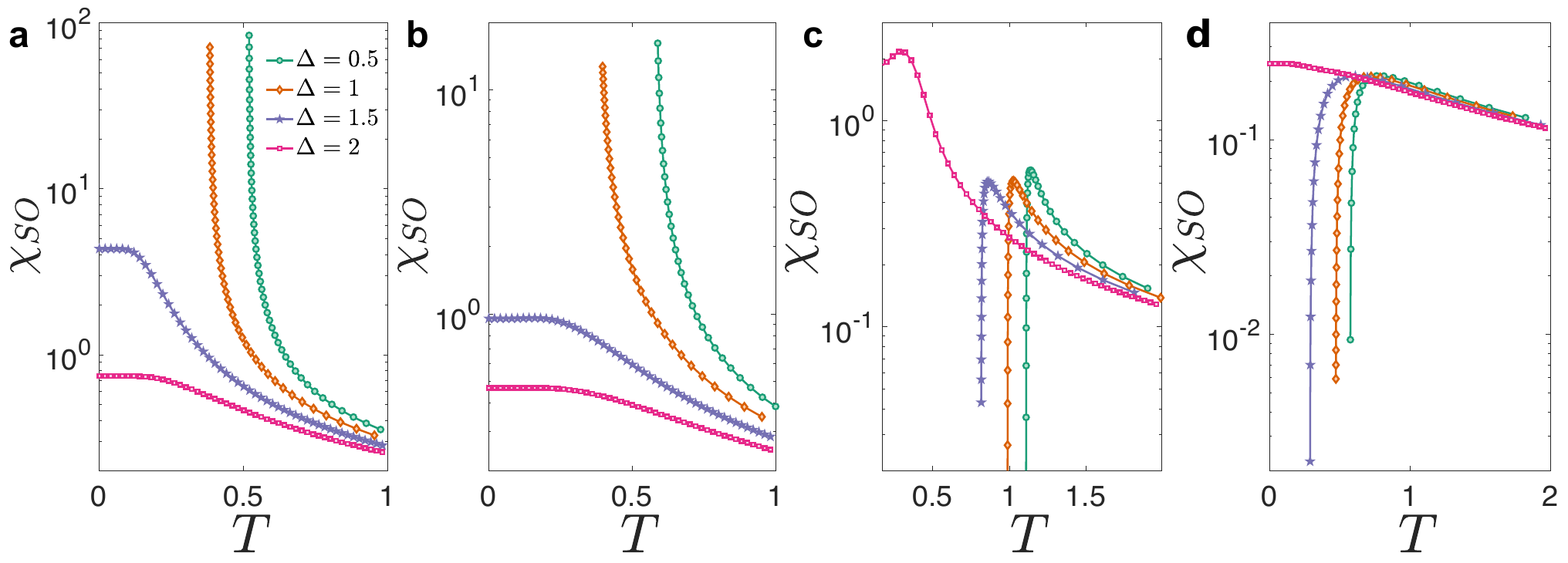}}
\caption{
{\bf Divergent susceptibility for a spin-orbit intertwined order with renormalization group calculation.}  The plots correspond to different choices of  interaction strengths ($U$, $J$, $J'$) and band gap ($\Delta$) in the tight binding model~\cite{supplement}. The tunneling of the $p$-orbital electron, or one half of bandwidth of the $p$-band is set as the energy unit here. In ({\bf a}, {\bf b}, {\bf c}, {\bf d}), we have $(U, J, J') = (2, 0, 0)$, $(U, J, J') = (2, -1, -0.5)$, $(2, 0.1, -0.5)$, and $(2, -1, 0.5)$, respectively, and set the $s$-band tunneling $t_s = 2$. With both of $J$ and $J'$ negligible (set to zero in this plot) in ({\bf a}),  a divergent susceptibility at finite temperature ($T$) in the time-reversal symmetric spin-orbit intertwined channel (Eq.~\eqref{eq:SOorder}) is confirmed, which agrees with the analytical results from random phase approximation. In ({\bf b}), we find that with both of $J$ and $J'$ being negative, the susceptibility in the spin-orbit intertwined channel is more strengthened, compared to ({\bf a}). In ({\bf c}), we find that a positive Hund's rule coupling $J>0$ does not immediately cause  suppression of the spin-orbit intertwined order. ({\bf d}) shows that a positive Josephson coupling $J' >0$ does not support the time-reversal symmetric spin-orbit intertwined order.}
\label{fig:f2}
\end{figure*}

In order to generate effective SOI with no spin polarization, we shall consider a spin-orbit intertwined order that breaks spin SU(2) symmetry and preserves time reversal symmetry.  From Table~\ref{tab:t1}, the potential candidates are, 
\be 
\hat{\cal O} _{m_s}  \equiv \frac{1}{\sqrt{2}} \left[ {\cal B}_{1, m_s, -1; 0 } -{\cal B} _{1, m_s, -1; 1} \right], 
\label{eq:SOorder} 
\ee 
and 
\be 
\hat{\cal O}' _{m_s }  \equiv \frac{1}{\sqrt{2}} \left[ {\cal B}_{1, m_s, 0; 1 } -{\cal B} _{1, m_s, 0; -1} \right], 
\label{eq:SOorder} 
\ee 
whose parities are odd and even, respectively. They both satisfy the symmetry requirement, but the parity even operators represent parings within $s$ or $p$-bands, which do not benefit from the divergent susceptibility in our model.  We thus focus on analyzing the parity odd spin-orbit intertwined operators $\hat{\cal O} _{m_s}$. 
We emphasize here that this symmetry channel would not be possible in a single-orbital model, where a local triplet order necessarily breaks time-reversal symmetry. 

Considering a weak Josephson coupling $J' $,  the susceptibility  towards forming an order of $\hat{\cal O} _{m_s}$
is, 
\be 
\chi_{\rm SO}  = \frac{\chi_{sp} ^0}{1-(U-J) \chi_{sp} ^0}
\label{eq:chiSO} 
\ee 
under  random phase approximation~\cite{supplement}. 
The divergence in the susceptibility for the non-interacting model at zero temperature persists to finite band gap and finite temperature, provided that $U>J$. This requirement is satisfied considering a typical situation for electrons in a molecule, that the  density-density interaction is repulsive and the Hund's rule coupling is ferromagnetic ($J<0$ in our notation). 
Without Josephson coupling, i.e., $J'=0$, a degenerate channel---$\left[ {\cal B}_{1, m_s, -1; 0 } + {\cal B} _{1, m_s, -1; 1} \right]/\sqrt{2}$ which would break time reversal symmetry yields a susceptibility of an identical strength.  A negative Josephson coupling $J'<0$  would break this degeneracy, and  make the time reversal symmetric pairing  more favorable. 
In the case of Josephson coupling being negligible, as is expected for elongated chiral molecules (see Supplmentary Material), 
the presence of orbital motion induced Zeeman splitting selects the time reversal symmetric over the asymmetric  pairing.

Since all parity odd pairing channels in our theory potentially have a large susceptibility due to the divergence in $\chi_{sp} ^0$, we further  go beyond the random phase approximation, and carry out a systematic submission of Feynman diagrams using a scheme of renormalization group flow equation, which incorporates the intertwined scatterings among different channels~\cite{2012_Metzner_RMP}. The results for the susceptibility of forming the spin-orbit intertwined order in Eq.~\ref{eq:SOorder} are shown in Fig.~\ref{fig:f2}. It is confirmed that having $J$ and $J'$ vanishing, the susceptibility for $\hat{\cal O}_{m_s}$ diverges at finite temperature (Fig.~\ref{fig:f2}{\bf a}). The chiral peptide $3_{10}$ helix corresponds to this case according to interaction strength estimate~\cite{supplement}. The divergence requires the interaction energy to conquer  the band gap barrier of the $sp$-orbital pairing. 
The susceptibility is more strengthened for both $J$ and $J'$ being negative (Fig.~\ref{fig:f2}{\bf b}). 
We also find that having a weak antiferromagnetic coupling in $J$ still supports divergent susceptibility in $\hat{\cal O}_{m_s}$ channel (Fig.~\ref{fig:f2}{\bf c}). 
For a large enough positive $J$, this susceptibility is no longer divergent, but instead meets a strong suppression at low temperature, which is due to a divergent susceptibility in a different channel, $\CB_{0, 0, \pm 1, q} $~\cite{supplement}. 
A sign change in the Josephson coupling leads to a low-temperature suppression of $\hat{\cal O}_{m_s}$ pairing (Fig.~\ref{fig:f2} {\bf d}), owing to a divergent susceptibility in the time-reversal odd channel of $\left[ {\cal B}_{1, m_s, -1; 0 } + {\cal B} _{1, m_s, -1; 1} \right]/\sqrt{2}$~\cite{supplement}. 
We expect that further considering scattering with additional orbitals present in molecules would renormalize $J'$ to the negative side due to BCS channels~\cite{2012_Metzner_RMP}, which would make the time-reversal symmetric spin-orbit intertwined order even more favorable.

A divergence in the susceptibility $\chi_{\rm SO}$ implies the correlation length of $\langle \hat{\cal O}_{m_s} ^\dag (z) \hat{\cal O}_{m_s'} (z')  \rangle$ reaches the size of the molecule. Taking mean field approximation, the operator $\hat{\cal O}_{m_s}$ acquires a finite expectation value, $\langle \hat{\cal O}_{m_s}\rangle \equiv \Phi_{m_s}$. The free energy of the order parameter takes an SU(3) symmetric form as shown in Supplementary Material. The pairings with different $m_s$ quantum numbers $0, \pm 1$ are degenerate,   
although they lead to two distinctive many-body states, analogous to the spin-$1$ polar and ferromagnetic superfluids in spinor condensate~\cite{1998_Ohmi_JPSJ,1998_Ho_PRL} or liquid Helium~\cite{volovik2003universe}. 

Further considering a circular motion induced Zeeman splitting for electrons~\cite{2015_Naaman_Review}, we have a perturbative coupling 
$ 
\Delta H = \delta\int dz 
\left[ i\psi_{x\uparrow} ^\dag    \psi_{y\uparrow} - i\psi_{x\downarrow} ^\dag   \psi_{y\downarrow}   + H.c. \right]$.   
Despite its insufficient strength to model CISS~\cite{2015_Naaman_Review,2018_Rebergen_CIIS}, this term determines a spin quantization axis, and triggers an order $(\Phi_{+1} \equiv \phi , \Phi_0 = 0, \Phi_{-1} = 0)$ to minimize to total free energy in our theory, from which a strong SOI emerges from electron correlation, 
\be 
\textstyle H_{\rm SOI} = \frac{U-J-J'}{4 \sqrt{2} } 
\int dz \textstyle \left[ i \phi \Psi^\dag (z) 
	  \sigma_+ \otimes L_- \Psi(z) + H.c. \right],  
\label{eq:HSOI} 
\ee 
with $\Psi\equiv [\psi_{1, \up} , \psi_{0, \up} \psi_{-1, \up}, \psi_{1, \down} , \psi_{0, \down} \psi_{-1, \down}]^T$,  
$\sigma_+$ and $L_-$ the standard spin-$1/2$ and spin-$1$ angular momentum matrices~\cite{supplement}. 

This correlation induced coupling breaks spin SU(2) and reflection symmetries with the time reversal symmetry unbroken.  Through a unitary transformation into the quasi-particle basis, the induced coupling in the conduction band takes a more standard SOI form given in Eq.~\eqref{eq:SOIstrength}. 
{The induced SOI strength for the quasi-particles reaches to the order of the energy scale of room temperature, and is sufficient to model the spin dependent transmission observed in chiral molecule experiments~\cite{2018_Rebergen_CIIS}.}  

We remark here that Mermin-Wegner theorem does not forbid the long-range order formation in our setup as the continuous symmetries of spatial rotation and the spin SU(2) are all weakly broken considering the real geometry of a chiral molecule and the circular motion induced Zeeman splitting.


{\it Discussion.---} 
We have developed a novel quantum mechanism for strong SOI to emerge from many-body correlation effect. This  provides  an alternative origin for SOI, other than the fundamental quantum electrodynamics, which is particularly crucial to the understanding of the large CISS observed in  chiral organic molecules, where the bare SOI is too weak for modeling the experimental observation. Our theory may provide important guidance for future searching of other chiral molecules with CISS, and potentially contributes to the fundamental understanding of spin-selective biological processes. 

Since SOI plays an important role in topological physics in general, we expect the mechanism of correlation induced SOI may also shed light on engineering of topological devices such as Majorana quantum computing qubits and also neutral-atom based quantum simulations of topological physics where the bare SOI is absent. 

{\bf Acknowledgements.}
We acknowledge helpful discussion with Martin Plenio, Mikhail Lemeshko, Gang Chen, Hongjun Xiang, and Yinghai Wu. 
This work is supported by National Program on Key Basic Research Project of China (Grant No. 2017YFA0304204), National Natural Science Foundation of China (Grants No. 11934002 and 11774067), Natural Science Foundation of Shanghai City (Grant No. 19ZR1471500), Shanghai Municipal Science and Technology Major Project (Grant No. 2019SHZDZX01).

\bibliography{references}
\bibliographystyle{naturemag}

\begin{widetext}

\newpage
\renewcommand{\theequation}{S\arabic{equation}}
\renewcommand{\thesection}{S-\arabic{section}}
\renewcommand{\thefigure}{S\arabic{figure}}
\renewcommand{\thetable}{S\arabic{table}}
\setcounter{equation}{0}
\setcounter{figure}{0}
\setcounter{table}{0}

\begin{center} 
{\Huge Supplementary Material} \\
\end{center}

\section{Tight binding model}

The tight binding Hamiltonian corresponding to the field theory in Eq.~(3) is 
\bea 
H_0 &=& \frac{\Delta}{2} \sum_{j\alpha}  \left[ c_{p_x \alpha, j} ^\dag c_{p_x \alpha, j} + c_{p_y \alpha, j} ^\dag c_{p_y \alpha, j} - c_{s \alpha, j} ^\dag c_{s \alpha, j} \right ] \nn   \\
&+&t_s  \sum_{j\alpha}  \left[ c_{s \alpha,j} ^\dag c_{s \alpha, j+1} +  c_{s \alpha,j} ^\dag c_{s \alpha, j-1} -2 c_{s \alpha,j} ^\dag c_{s \alpha, j}\right] \\
&-&t_p  \sum_{j\alpha, \nu = p_x, p_y }  \left[ c_{\nu \alpha,j} ^\dag c_{\nu \alpha, j+1} +  c_{\nu \alpha,j} ^\dag c_{\nu \alpha, j-1} -2 c_{\nu \alpha,j} ^\dag c_{\nu \alpha, j} \right],  \nn 
\eea  
with $j$ the site index of the tight binding model, $t_s$ and $t_p$ the nearest neighboring tunneling of the $s$- and $p$- orbitals, and $c_{\nu \alpha}$ the lattice annihilation operator associated with $\psi_{\nu \alpha}$. The energy dispersion of the tight binding model 
is $\epsilon_\nu (k)  = 2 t_\nu (1-\cos k ) P_\nu $. 
The tunneling parameters relate to the effective mass in the field theory as $m_\nu ^{-1} = 2 t_\nu/\hbar^2$, taking the lattice constant as a length unit. 

Taking spin SU(2) and rotation symmetries, the local electron interaction between $s$- and $p$-orbitals involves a 
density-density term, 
\be  
\textstyle U\sum_j   C^\dag_{\nu,j} C_{\nu,j}   C_{s,j} ^\dag C_{s, j}  
\label{eq:Usp} 
\ee 
the Hund's rule coupling, 
\be 
J \sum_{\nu = p_x, p_y}  C^\dag_{\nu, j} \vec{\sigma} C_{\nu, j}  \cdot C_{s,j}  ^\dag \vec{\sigma}  C_{s, j}  ,   
\label{eq:Jsp} 
\ee 
and a Josephson coupling, 
\be 
 J' \sum_j \left[ \left( c_{p_x\uparrow, j}^\dag  c_{p_x \downarrow , j} ^\dag  +   c_{p_y\uparrow, j}^\dag  c_{p_y \downarrow , j} ^\dag \right)  
 c_{s \downarrow, j} c_{s \uparrow, j} + H.c. \right] , 
 \label{eq:Jspprim} 
\ee 
where we have introduced compact notation, $C_{\nu, j} \equiv [c_{\nu, \uparrow, j} , c_{\nu, \downarrow, j} ]^T$.  
The intra-orbital interaction within the valence band or the conduction band is not included here because only the inter-band interactions  contribute strongly to the divergent susceptibility considering the divergence in $\chi_{sp} ^0$.

\section{Renormalization group flow}

In the calculation of susceptibility, we use the scheme of renormalization group flow to carry out a systematic resummation  of higher order Feynman diagrams, in order to capture the complex  intertwined scatterings among different channels in our theory. Having SU(2) symmetry, the one particle irreducible (1PI) four point function 
\be 
 \Gamma_{\nu_1 \alpha_1, \nu_2 \alpha_2; \nu_3 \alpha_3, \nu_4 \alpha_4} 
= \langle  \psi_{\nu_1 \alpha_1}  
 \psi_{\nu_2 \alpha_2}   \psi_{\nu_3 \alpha_3} ^\dag  \psi_{\nu_4 \alpha_4} ^\dag 
 \rangle_{\rm 1PI}, 
\ee 
takes a restricted form 
\bea 
&& \Gamma_{\nu_1 \alpha_1, \nu_2 \alpha_2; \nu_3 \alpha_3, \nu_4 \alpha_4}     \\
&=& V_{\nu_1 \nu_2 \nu_3 \nu_4} \delta_{\alpha_1 \alpha_4} \delta_{\alpha_2 \alpha_3} 
-V_{\nu_2 \nu_1 \nu_3 \nu_4}  \delta_{\alpha_1 \alpha_3} \delta_{\alpha_2 \alpha_4}  \nn 
\eea  
This function takes real values according to the time-reversal symmetry. 
Further considering rotation symmetry, the nonzero ones are 
$V_{ssss}$, 
$V_{xxxx}=V_{yyyy}$, 
$V_{ssxx}  = V_{ssyy} = V_{xxss} = V_{yyss}$, 
$V_{sxsx} = V_{sysy} = V_{xsxs} = V_{ysys}$, 
$V_{xssx} = V_{yssy} = V_{sxxs} = V_{syys}$, 
$V_{xxyy} = V_{yyxx}$, 
$V_{xyxy} = V_{yxyx}$, 
$V_{yxxy} = V_{xyyx}$. 
These four point functions are obtained by solving a renormalization group flow equation~\cite{2012_Metzner_RMP}, 
\bea 
 && \partial_l V(\nu_1, \nu_2, \nu_3, \nu_4) \nn \\ 
&&= \sum_{\nu \nu'} 
\left\{
-\dot{\Pi}_{\nu \nu'} ^{\rm pp} V_{\nu_1 \nu_2 \nu\nu'} V_{\nu \nu'\nu_3 \nu_4} 
+ 2 \dot{\Pi}_{\nu\nu'} ^{\rm ph}   V_{\nu_1 \nu' \nu \nu_4} V_{\nu \nu_2 \nu_3 \nu'} 
\right.  \\ 
&&\left.
- \dot{\Pi}_{\nu\nu'} ^{\rm ph} 
					\left[
					V_{\nu_1 \nu \nu_3 \nu'} V_{\nu'\nu_2 \nu \nu_4} 
					+V_{\nu \nu_2 \nu_3 \nu'} V_{\nu' \nu_1 \nu \nu_4} 
					+V_{\nu_1 \nu'\nu \nu_4} V_{\nu_2 \nu \nu_3 \nu'} 
					\right] 
\right \}. \nn
\label{eq:fullflow} 
\eea 
Here  $l$ is a parameter running from $0$ from $+\infty$, and $\dot{\Pi}$ represents derivatives $\Lambda\partial_\Lambda$ of the functions, 
\[
{\Pi}_{\nu \nu'} ^{\rm pp} (\Lambda)  = \int \frac{dk}{2\pi} \frac{n_f(-\epsilon_\nu(k)) -n_f (\epsilon_{\nu'} (k) )}{\epsilon_\nu (k) + \epsilon_{\nu'} (k)}  
[\Theta_< (\epsilon_\nu (k) ) \Theta_<(\epsilon_{\nu'} (k)) ]
\] 
\[ 
{\Pi}_{\nu \nu'} ^{\rm ph} (\Lambda)  = \int \frac{dk}{2\pi} \frac{n_f(\epsilon_\nu(k)) -n_f (\epsilon_{\nu'} (k) )}{\epsilon_\nu (k) - \epsilon_{\nu'} (k)}  
[\Theta_< (\epsilon_\nu (k) ) \Theta_<(\epsilon_{\nu'} (k)) ] 
\] 
where $n_f (\epsilon)$ is the Fermi-Dirac distribution function, and
 $\Theta_< (\epsilon) =  \frac{|\epsilon|/[\Lambda] }{e^{|\epsilon|/[\Lambda] } -1}$, 
 with $\Lambda = \Lambda_0 e^{-l}$, 
introduced to continuously integrate out high energy modes in the renormalization group flow. 
 For the  initial condition at $l = 0$, $\Lambda_0$ is set to be much larger than the bandwidth, and the four point functions are initialized as  $V_{ssxx}  = V_{ssyy} = V_{xxss} = V_{yyss} = J'$, $V_{sxsx} = V_{sysy} = V_{xsxs} = V_{ysys} = -2 J$, $V_{xssx} = V_{yssy} = V_{sxxs} = V_{syys} = U-J$, with others initialized at $0$ in absence of intra-band interaction. 
The generated intra-band interactions in the renormalization group flow are kept in our calculation. 


\section{Calculation of susceptibility}

In presence of an external perturbation 
$\delta H = \int dz  \sum_{\nu \alpha  \nu' \alpha'} h_{\nu \alpha , \nu' \alpha'} \psi_{\nu  \alpha} ^\dag \psi_{\nu'\alpha'} $, 
the susceptibility is obtained from linear response theory, 
\bea  
\label{eq:chifull} 
&& \chi_{\nu_1 \alpha_1, \nu_2 \alpha_2; \nu_3 \alpha_3, \nu_4 \alpha_4 }  
=\frac{\partial \langle \psi_{\nu_2 \alpha_2} ^\dag \psi_{\nu_3 \alpha_3}  \rangle }{\partial h_{\nu_1 \alpha_1, \nu_4 \alpha_4} }  \nn \\ 
&=& -\delta_{\sigma_1 \sigma_3 } \delta_{\sigma_2 \sigma_4 } \delta_{\nu_1 \nu_3}  \delta_{\nu_2 \nu_4} \Pi_{\nu_1 \nu_2} ^{\rm ph} (\Lambda_0)  \\ 
&-& \Gamma_{\nu_1 \sigma_1, \nu_2 \sigma_2; \nu_3 \sigma_3, \nu_4 \sigma_4} \Pi_{\nu_1 \nu_4} ^{{\rm ph}} (\Lambda_0) \Pi ^{\rm ph} _{\nu_2 \nu_3} (\Lambda_0) \nn 
\eea  
where the self-energy corrections are neglected following a standard approximation in simplification of  functional renormalization group flow~\cite{2012_Metzner_RMP}.
The first term in Eq.~\eqref{eq:chifull} is the non-interacting part, and the specification to $\nu_1  = \nu_3 = s$ and $\nu_{2} = \nu_4 = p_x$ (or $p_y$) leads to the expression of $\chi_{sp} ^0$ at the zero temperature limit. 
The susceptibilities in the channels of $\CB _{j_s, m_s, m_l; q}$ (Eq.~(4)) have a block diagonal form, 
\bea 
&& \chi_{j_s m_s m_l } (q, q') = \\ 
& \sum & 
U_{j_s m_s m_l q; \nu_4 \sigma_4, \nu_1 \sigma_1 } 
U^*_{j_s m_s m_l q'; \nu_2 \sigma_2, \nu_3 \sigma_3} 
\chi_{\nu_1 \sigma_1, \nu_2 \sigma_2; \nu_3 \sigma_3, \nu_4 \sigma_4} \nn
\eea 
The symbol $\sum$ performs summation over the indices $\nu_1 \sigma_1$, $\nu_2 \sigma_2$, $\nu_3 \sigma_3$, and $\nu_4 \sigma_4$.  
The unitary transformation is determined according to the definition of $\CB$ operators as, 
\be 
U_{j_s m_s m_l q; \nu \sigma, \nu'\sigma'} = (-)^{q+1} i^{2\sigma } C_{\frac{1}{2} \frac{1}{2}} (-\sigma, \sigma'| j_s m_s) T_{q \nu} ^* T_{q + m_l, \nu'} 
\ee 
with $T$ the matrix corresponds to  transformation of $s$- and $p$-orbitals into the angular momentum basis (Eq.~(4)). 
The susceptibility  towards the order formation in $\hat{O}_{m_s}$ channels is obtained as  
\be 
\chi_{\rm SO} = \chi_{1 m_s ,-1}(0, 0) - \chi_{1 m_s, -1} (0, 1) = -\Pi_{sp} ^{\rm ph} (\Lambda_0) - [V_{sxxs} (l\to+ \infty)  -V_{xxss} (l\to+\infty)  ][\Pi_{sp} ^{\rm ph} (\Lambda_0)]^2 .  
\ee 
The numerical results shown in the main text are obtained by solving the full flow equation. 
The analytic expression in Eq.~(7)  is obtained in the framework of  renormalization group flow by solving the flow equation keeping the divergent ($\sim \Delta^{-1/2} $) contributions only, yielding 
$ 
\partial_l V_{xssx} =- \dot{\Pi}^{\rm ph} _{sp}  V_{xssx} ^2 .
$

\section{Free energy of the spin-orbit intertwined order}
 
Having a diverging correlation length in $\langle \hat{\cal O}_{m_s} ^\dag (z) \hat{\cal O}_{m_s'} (z')  \rangle$ implies the operator acquires  an expectation value $\langle \hat{\cal O}_{m_s } \rangle = \Phi_{m_s} $, which is a three-component complex field. From SU(2) symmetry, the associated Ginzburg Landau free energy reads as 
\be 
F = -r  \Phi^\dag \Phi  + c_0 ( \Phi^\dag \Phi )^2 + c_2 (\Phi^\dag \vec{L} \Phi)\cdot (\Phi^\dag \vec{L} \Phi) + {O} (|\Phi| ^6) , 
\ee 
with $Lx$, $L_y$, $L_z$ the standard spin-$1$ representation matrix of SU(2) group. 
To relate to our Fermionic theory, the phenomenological couplings $r$, $c_0$ and $c_2$ are calculated with a Hubbard-Stratonovich transformation as 
\bea 
r &=& [J+J' -U] + \int \frac{dk}{2\pi} \frac{[J+J' -U]^2 }{\epsilon_p(k) - \epsilon_s (k) } , \nn \\
c_0 &=&  \frac{1}{2} \int \frac{dk}{2\pi} \frac{[J+J' -U]^4}{[\epsilon_p(k) - \epsilon_s (k)]^3}. 
\eea 
In our theory, the coefficient $c_2$ vanishes for  an accidental  symmetry, causing an emergent SU(3) symmetry in the free energy.  
Then all states with the same order parameter amplitude $\sqrt{ \Phi^\dag \Phi}$ are degenerate in energy, although there are two distinctive states analogous to the polar and ferromagnetic phases in spin-$1$ superfluids~\cite{1998_Ho_PRL,volovik2003universe}.  
The order parameter strength is then given by $\sqrt{ \frac{r}{2c_0} }$, and its small band gap limit is given in the main text. 


However the above degeneracy is lifted by considering  a circular motion induced Zeeman splitting for electrons, which is described by a Hamiltonian 
\be 
\Delta H = \delta\int dz 
\left[ i\psihat_{x\uparrow} ^\dag    \psihat_{y\uparrow} - i\psihat_{x\downarrow} ^\dag   \psihat_{y\downarrow}   + H.c. \right].
\ee 
This leads to a symmetry-breaking term in the Free energy, 
\be 
\Delta F  = -\delta \sum_{m_s } m_s | \Phi_{m_s} |^2  \int \frac{dk}{2\pi} [\epsilon_p(k) - \epsilon_s(k) ]^{-2}. 
\ee  
Minimizing the total free energy leads to $\Phi_+ = \phi$, $\Phi_0 = \Phi_- =0$ for $\delta >0$. The remaining unbroken $U(1)$ symmetry, $\phi \to \phi e^{i\theta}$, corresponds to the rotation symmetry around the $z$ direction.

\section{Quasi-particle Hamiltonian}
 
Having  an order $\Phi_+ = \phi$, $\Phi_0 = \Phi_- =0$, the bare electrons turn into dressed quasi-particles, whose Hamiltonian reads as 
$H_{\rm QP} = H_0 + H_{\rm SOI} $, with 
\be 
H_{\rm SOI}  = -(U-J-J') \int dz \left[ \phi^* \hat{\cal O}_{+1} (z) + H.c. \right]. 
\ee 
Its explicit form in terms of $\psihat$ operators is given in Eq.~(8), where we have invoked  spin-$1/2$ and spin-$1$ matrices 
\be 
\sigma_+  =\left[ 
	\begin{array}{cc} 
	0 &2 \\
	0 & 0 
	\end{array} 
	 \right] , 
L_- =\left[ 
	\begin{array}{ccc} 
	0	&0 	&0 \\
	2	&0	&0 \\ 
	0	&2	&0 
	\end{array} 
	\right]. 
\ee 

The eigenmodes describing quasi-particles are introduced through a unitary transformation, 
\be
\left[ 
\begin{array}{c} 
\tilde{\psihat}_{+\up } (k) \\
\tilde{\psihat}_{0\down} (k)  \\ 
\tilde{\psihat}_{-\up}   (k) 
\end{array} 
\right ] 
= 
\left[ 
\begin{array}{ccc} 
1	&0							&0\\ 
0	&\cos (\vartheta_k/2) e^{i\varphi/2} 	&-\sin(\vartheta_k/2) e^{-i\varphi/2} \\ 
0	&\sin(\vartheta_k /2) e^{i\varphi/2} 	&\cos(\vartheta_k/2) 	e^{-i\varphi/2} 
\end{array} 
\right] 
\left[ 
\begin{array}{c} 
{\psihat}_{+\up } (k) \\
{\psihat}_{0\down} (k)  \\ 
{\psihat}_{-\up} (k)    
\end{array} 
\right ], 
\ee 
\be
\left[ 
\begin{array}{c} 
\tilde{\psihat}_{+\down } (k) \\
\tilde{\psihat}_{0\up} (k)  \\ 
\tilde{\psihat}_{-\down}   (k) 
\end{array} 
\right ] 
= 
\left[ 
\begin{array}{ccc} 
\cos(\vartheta_k/2) 	e^{-i\varphi/2} 	&\sin(\vartheta_k /2) e^{i\varphi/2} 		&0\\ 
-\sin(\vartheta_k/2) e^{-i\varphi/2} 		&\cos (\vartheta_k/2) e^{i\varphi/2} 	&0\\ 
0								&0	 							&1 
\end{array} 
\right] 
\left[ 
\begin{array}{c} 
{\psihat}_{+\down } (k) \\
{\psihat}_{0\up} (k)  \\ 
{\psihat}_{-\down} (k)    
\end{array} 
\right ], 
\ee 
with 
\[
\cos \vartheta_k = [\epsilon_p (k) - \epsilon_s (k) ]/\sqrt{ [\epsilon_p(k) - \epsilon_s (k) ]^2 + 2 (U-J-J') ^2 |\phi|^2 },  
\]   
\[
\sin\vartheta_k = (U-J-J') |\phi |/  \sqrt{ [\epsilon_p(k) - \epsilon_s (k) ]^2 + 2 (U-J-J') ^2 |\phi|^2 },
\] 
and 
$\varphi = \arg (-i\phi^*)$.   
We emphasize that the renormalized modes $\tilde{\psihat}_{m_l ,\alpha}$ still mainly carry their original spin moment and angular momentum. 

The induced quasi-particle Hamiltonian associated with the conduction band is then obtained to be 
\be 
H_{\rm SOI} = \int \frac{ dk }{2\pi}  \frac{ \lambda_{\rm so}  (k) } {2} 
\left[  
\tilde{ \psihat} _{+\uparrow} ^\dag (k)   \tilde{ \psihat} _{+\uparrow} (k)  
+\tilde{ \psihat} _{-\down} ^\dag (k)   \tilde{ \psihat} _{-\down} (k)  
- \tilde{ \psihat} _{-\uparrow} ^\dag (k)   \tilde{ \psihat} _{-\uparrow} (k) 
-\tilde{ \psihat} _{+\down} ^\dag (k)   \tilde{ \psihat} _{+\down} (k)  
\right], 
\ee 
with a momentum dependent SOI strength 
\[
\textstyle \lambda_{\rm so} (k_z) 
= [\epsilon_p (k_z) - \epsilon_s (k_z) ]/2 
 - \sqrt{ [\epsilon_s (k_z) - \epsilon_p (k_z) ]^2/4 + (J-U-J')^2 |\phi|^2 /2 } . 
\]  
Near the band edge, the induced SOI strength further simplifies to the expression given in Eq.~(1)

\begin{figure*}[htp]
\centerline{\includegraphics[angle=0,width=\textwidth]{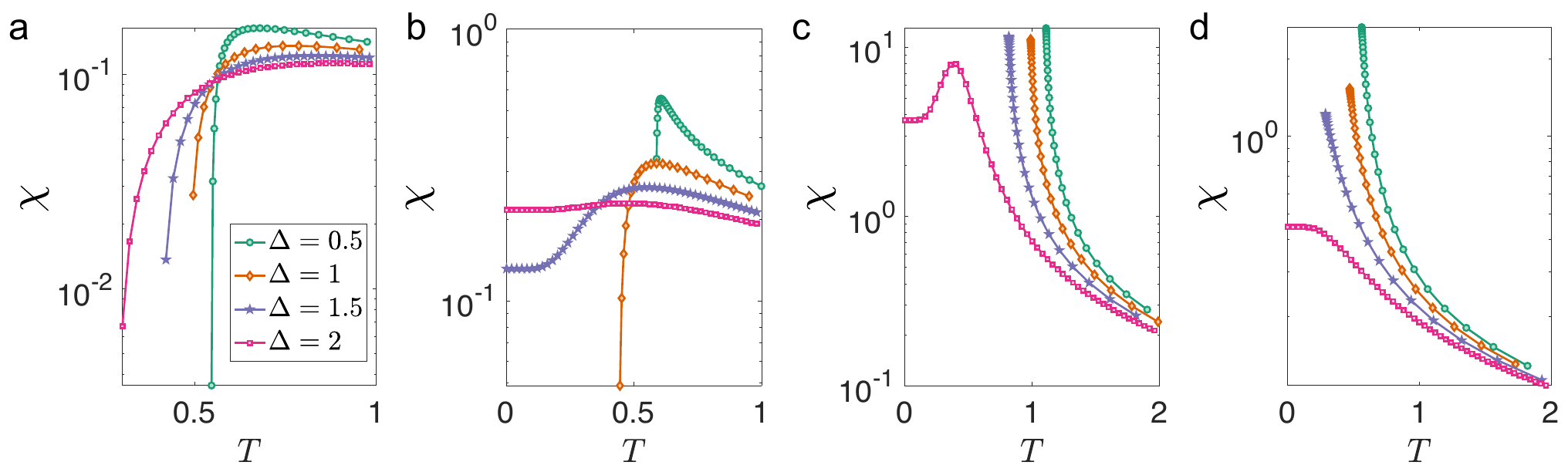}}
\caption{
{\bf The susceptibility corresponding to the channel of $ \left[{\cal B}_{0,0,1;0} - {\cal B}_{0,0,1;-1} \right] /\sqrt{2}$ with renormalization group calculation.}  The plots correspond to different choices of  interaction strengths ($U$, $J$, $J'$) and band gap ($\Delta$) in the tight binding model. The tunneling of the $s$-orbital electron, or one half of bandwidth of the $s$-band is set as the energy unit here. In ({\bf a}, {\bf b}, {\bf c}, {\bf d}), we have $(U, J, J') = (2, -1, -0.5)$, $(2, 0.1, -0.5)$, $(2, 1, -0.5)$, and $(2, -1, 0.5)$, respectively. This susceptibility remains non-divergent at low temperature in {\bf a}, and {\bf b}. In {\bf c}, and {\bf d}, we find a divergent susceptibility, which causes the strong suppression of spin-orbit intertwined order at low temperature.}
\label{fig:fs1}
\end{figure*}




\section{Symmetry properties} 
In this supplementary section, we provide more details of our symmetry analysis. 
In the basis of $\psihat_{q\alpha}$, the anti-unitary  time-reversal symmetry transformation (${\cal T}$) is represented as~\cite{Weinberg1995TheQT}, 
\bea 
{\cal T} \psihat_{q\alpha} {\cal T} ^{-1}  = i^{-2\alpha} (-1)^q \psihat_{-q, -\alpha}, \\ 
{\cal T} \psihat_{q\alpha} ^\dag {\cal T} ^{-1}  = i^{2\alpha} (-1)^q \psihat_{-q, -\alpha}.  \\ 
\eea 
The spatial parity transformation (${\cal P}$) is represented as 
\bea 
{\cal P} \psihat_{q \alpha} {\cal P}^\dag = (-1)^q \psihat_{q, \alpha}. 
\eea 
Under spatial rotation ($R_\theta $) around the $z$-axis, we have 
\be 
R_\theta \psihat_{q \alpha} R_\theta ^\dag  = \psihat_{q} e^{iq\theta}
\ee 
Then the corresponding symmetry properties of the composite operators ${\cal B}_{j_s, m_s, m_l; q}$ are derived as listed in Table~I in the main text. 
\medskip

\begin{figure*}[htp]
\centerline{\includegraphics[angle=0,width=\textwidth]{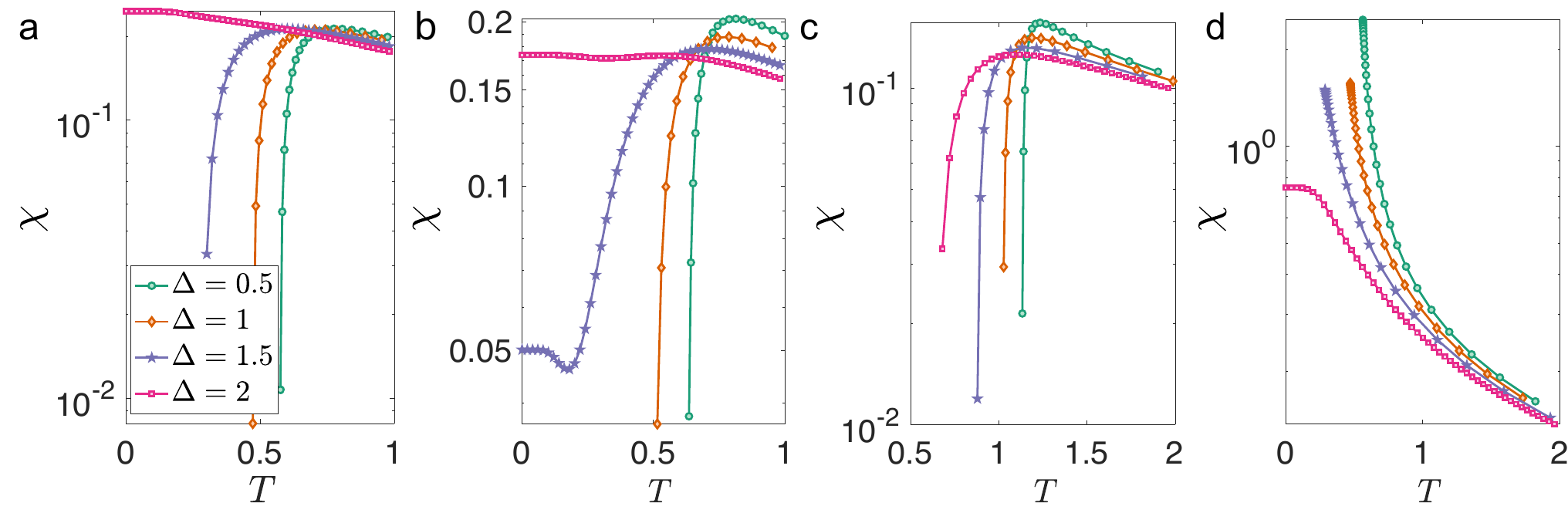}}
\caption{
{\bf The susceptibility corresponding to the channel of $\left[ {\cal B}_{1, m_s, -1; 0 } + {\cal B} _{1, m_s, -1; 1} \right]/\sqrt{2}$.}  
The parameter choices are the same as in Fig.~\ref{fig:fs1}. This susceptibility remains non-divergent at low temperature in {\bf a}, {\bf b} and {\bf c}. In {\bf d}, we find  the susceptibility diverges. 
}
\label{fig:fs2}
\end{figure*}

\section{Susceptibilities in other channels} 
In this supplementary section, we provide other relevant susceptibility channels which affect the spin-orbit intertwined order in the functional renormalization group flow. 
In Fig.~\ref{fig:fs1}, we provide the susceptibility corresponding to a time-reversal odd  spin singlet channel, 
\be 
\left[{\cal B}_{0,0,1;0} - {\cal B}_{0,0,1;-1} \right] /\sqrt{2}. 
\ee 
In Fig.~\ref{fig:fs1}({\bf a}, {\bf b}), this susceptibility does not diverge at low temperature, which then does not cause suppression of the spin-orbit intertwined order as shown in Fig.~2({\bf a}, {\bf b}) in the main text. In Fig.~\ref{fig:fs1}({\bf c}, {\bf d}), the susceptibility diverges at low temperature, which causes the strong suppression of the spin-orbit intertwined order at low temperature as shown in Fig.~2({\bf c}, {\bf d}).

In Fig.~\ref{fig:fs2}, we provide the susceptibility corresponding to a time-reversal odd spin triplet channel, 
\be 
\left[ {\cal B}_{1, m_s, -1; 0 } + {\cal B} _{1, m_s, -1; 1} \right]/\sqrt{2}. 
\ee 
This susceptibility is non-divergent in Fig.~\ref{fig:fs2}({\bf a}, {\bf b}, {\bf c}). In Fig.~\ref{fig:fs2}({\bf d}), this susceptibility diverges.

\begin{figure*}[htp]
\centerline{\includegraphics[angle=0,width=.7\textwidth]{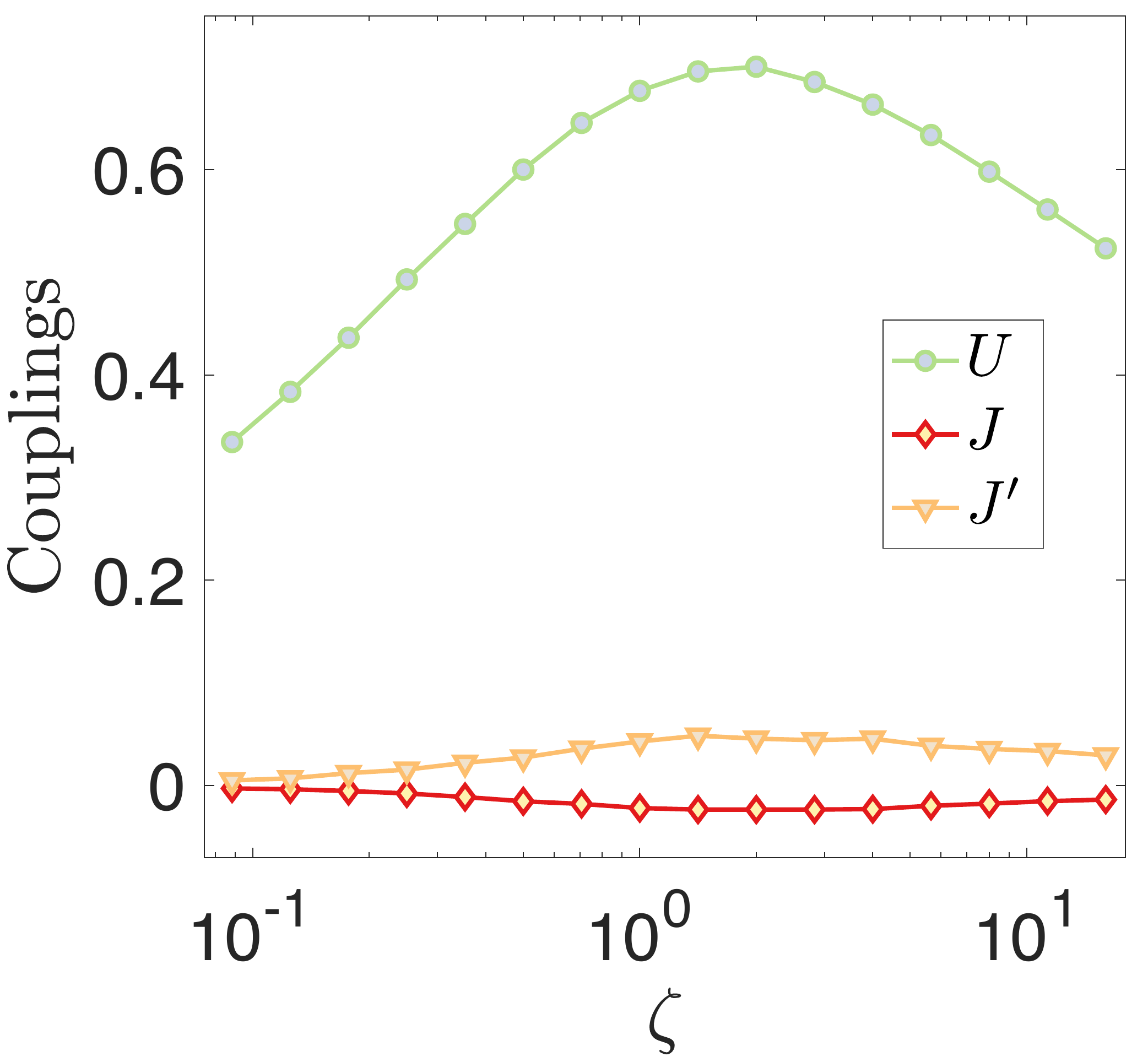}}
\caption{
{\bf The interaction strengths  in the tight-binding model.}  The energy unit in this plot is $E_0\equiv \frac{e^2}{\sqrt{a_\perp a_\parallel}}$, and the varied parameter $\zeta$ is a ratio $a_\parallel/a_\perp$ (see the description in Sec.~\ref{sec:IntEstimate}.) 
}
\label{fig:fs3}
\end{figure*}

\section{Estimate of interaction strengths} 
\label{sec:IntEstimate} 
Taking the {\it ab initio} band structure of a right-banded peptide helix~\cite{2020_Yan_CISS}, the low-energy properties of the molecule are captured by a three-orbital model having $p_x$, $p_y$, and $p_z$ orbitals---the $p_z$ orbital here is analogous to the $s$-orbital in our theory as they obey the same rotation symmetry. We estimate the interaction strength by taking a harmonic approximation to the Wannier functions of these orbitals ($\phi_{x,y,z} (\vec{r})$~\cite{2012_Vanderbilt_RMP}, 
\be 
\phi_{x,y,z} (\vec{r} ) \propto r_{x,y,z}   \exp\left\{ -\frac{1}{2} \sqrt{ \left( \frac{r_x} {a_\perp}  \right)^2 + 
								   \left( \frac{r_y} {a_\perp}  \right)^2 + 
								   \left( \frac{r_z}{a_\parallel} \right)^2 } \right\} , 					  
\ee
where $r_x$, $r_y$, $r_z$ are the three spatial coordinates, $a_\perp$ corresponds to radius of the molecule in the transverse direction, and $a_\parallel$ one-half of the size of the repetition unit along the molecular elongation direction. 
The bare interaction strengths at the starting point of the renormalization group flow are estimated by projecting the Coulomb interaction to the three $p$-orbitals, from which we have 
\be 
V_{\nu_1, \nu_2, \nu_3, \nu_4}/2 = \int d^3 \vec{r}  d^3 \vec{r}' \phi_{\nu_1} ^* (\vec{r}) \phi_{\nu_4} (\vec{r}) \frac{e^2} {|\vec{r}-\vec{r}'|} \phi_{\nu_2} ^* (\vec{r}') \phi_{\nu_3} (\vec{r}'), 
\ee 
with $\frac{e^2} {|\vec{r}-\vec{r}'|}$ the Coulomb interaction between two electrons. The interaction strengths $U$, $J$, and $J'$ in Eqs.~(S2,S3,S4) are obtained from $U = V_{xzzx}+V_{zxzx}/2$, $J  =- V_{zxzx}/2$, and $J' = V_{zzxx}$.  In this calculation, the interaction strengths are determined by $E_0\equiv \frac{e^2}{\sqrt{a_\perp a_\parallel}}$, and the ratio $\zeta \equiv a_\parallel/a_\perp$. The results are shown in Fig.~\ref{fig:fs3}, where $E_0$ is set as an interaction energy unit. With a choice of $\sqrt{a_\perp a_\parallel} = 0.5$nm, $E_0$ is about $5$eV. For a wide range choice of the ratio $\zeta$, $U$ is about $1.5\sim 3.3$eV, and both $J$ and $J'$ are below $0.1$eV, an order of magnitude smaller than $U$.

\end{widetext}

\end{document}